# A rigorous two-dimensional model for the stripline ferromagnetic resonance response of metallic ferromagnetic films

*Z. Lin*[1,2] *and M. Kostylev*[1*]

1. *School of Physics, The University of Western Australia, Crawley 6009, Australia*
2. *University of Science and Technology of China, Hefei, China*

Abstract. In this work we constructed a two-dimensional numerical model for calculation of the stripline ferromagnetic resonance (FMR) response of metallic ferromagnetic films. We also conducted numerical calculations by using this software. The calculations demonstrated that the eddy current contribution to the FMR response decreases with a decrease in the stripline width. The most important manifestations of the conductivity (eddy current) effect are excitation of the higher-order standing spin waves across the film thickness in the materials for which the standing spin wave peaks would be absent in cavity FMR measurements and strong dependence of the off-resonance series conductance of the stripline on the stripline width. Whereas the contribution of the eddy currents to the stripline FMR response can be very significant, because wide striplines (100nm+) are conventionally used for the FMR measurements, it is negligible in the case of excitation of spin waves, just because very narrow stripline transducers (0.5-5micron wide) are required in order to excite spin waves in metallic ferromagnetic films in a noticeable frequency/applied field range.

## 1. Introduction

The microwave conductivity contribution to the stripline broadband ferromagnetic resonance (FMR) response of highly-conducting (metallic) magnetic multilayers and nanostructures of sub-skin-depth thicknesses has attracted significant attention in recent years [1-8]. It has been shown that these effects are important every time the microwave magnetic field is incident on only one of the two surfaces of a planar metallic material [3,9-11].

The geometry of a stripline ferromagnetic resonance experiment [12-15,4] is characterized by such single-surface incidence of the microwave magnetic field on the sample surface [3]. This experiment usually employs a macroscopic coplanar (CPW) or microstrip (MSL) stripline through which a microwave current flows (Fig. 1). A sample - a film or a nanostructure - sits on top of this line separated by an insulating spacer (not shown in the figure) in order to avoid an electrical contact of the sample with the stripline. The stripline with the sample is placed in a static magnetic field applied along the stripline. The microwave current in the stripline drives magnetization precession in the ferromagnetic material. The complex transmission coefficient S21 of the stripline is measured either as a function of microwave frequency $f$ for a given applied magnetic field $\mathbf{H}=\mathbf{e}_z H$ ("frequency resolved FMR"), or as a function of $H$ for given $f$ ("field-resolved FMR") to produce FMR traces. The FMR absorption by the material is seen as a deep in the Re(S21) vs. $H$ or $f$ trace.

In the stripline FMR traces the effect of eddy currents is seen as anomalously large absorption amplitudes of the higher-order standing spin wave modes [3]. For single-layer ferromagnetic metallic films with unpinned surface spins on both film surfaces the standing spin waves modes are not seen at all in the cavity FMR spectra for symmetry reasons [16]. However, as an existing theory shows, in the case of the stripline FMR they may develop amplitudes up to 1/3 of the amplitude of the fundamental peak [17]. For the multilayered films lacking inversion symmetry in the direction of the stack thickness this effect may be much more dramatic [4].

*Corresponding author: mikhail.kostylev@uwa.edu.au

M. Bailleul [1] considered the electrodynamics of a thin metallic non-magnetic layer with a sub-skin-depth thickness on top of a coplanar stripline. He showed that there are two types of shielding by microwave eddy currents for the stripline configuration – electric and magnetic. The electric shielding is due to *capacitive* coupling of the layer to both signal and ground lines of CPW. Because a microwave voltage is applied between the signal line and the ground lines, this coupling results in a microwave current flowing from the signal line to the ground lines, i.e. in the direction *perpendicular* to the CPW line. (Essentially, the metallic layer acts a short-circuiting plug.)

It is clear, that the Oersted field induced by this current is along **H** and hence cannot contribute to the FMR response of a metallic ferromagnetic layer. However, in addition to electric shielding magnetic shielding may simultaneously exist. This type of shielding is due to *inductive* coupling of the metallic layer to the signal line of the CPW. The inductive coupling results in a microwave current flowing in the layer in the direction *opposite* to the direction of the current in the signal line. The origin of the current in the metallic layer is the same as in the case of the incidence of a plane electromagnetic wave on a metal surface – an eddy current is induced in the material. The direction of the eddy current is such that the total microwave magnetic field is enhanced in front of the sample (i.e strong back-reflection of the electromagnetic wave takes place) and gradually drops inside the metal with a distance from its surface (skin effect).

For the metallic layers with sub-skin-depth thicknesses the skin effects takes an unusual form – the microwave magnetic field inside the layer drops *linearly* from a maximum at the surface exposed to the microwave radiation to *zero* at the far film surface. For the ferromagnetic metallic films this was first theoretically demonstrated by solving a simplified one-dimensional problem [3,11]. Later this effect was confirmed with more rigorous two-dimensional numerical modelling. The 2D study was carried out for *non-magnetic* films [1,18,17].

The simplified quasi-analytical 1-dimensional approach from [3] is valid for striplines whose width $w$ can be assumed as infinite – MSLs with large widths of the strip and CPWs with large widths of the signal line. The 2D calculations mentioned above assume realistic geometry of the stripline cross-section – they account for the finite widths of the strip of MSL and of the signal line of CPW as well as for the presence of the ground lines in the CPW case. It has been shown that the strength of shielding by the eddy currents strongly depends on $w$. The amplitude of the microwave magnetic field at the far film surface is practically zero for $w$ in the range 0.3 to 1.5mm. For smaller $w$-values it gradually decreases with $w$ and becomes almost equal to the field value at the front film surface for $w$=50 micron or so and for realistic values of all other parameters of the calculation [18,17].

In the present work we construct a numerical model and solve the 2-dimensional electromagnetic problem numerically for *ferromagnetic* films. In addition to the film conductivity and electromagnetic interactions, the exchange and Zeeman interactions are taken into account. We focus specifically on the case of the MSL lines. A simple estimation demonstrates that the electric current shielding is negligible for microstrip lines due to a large (electrical) distance between the sample and the ground plane of the MSL (Fig. 1) resulting in an almost negligible capactitive conductance between the film and the ground plane.

Our numerical results show that the FMR absorption amplitude for the 1st standing spin wave mode is large for strips of macroscopic sizes and gradually drops to zero with a decrease in $w$. This happens due to the decrease in the microwave shielding effect, as previously found for non-magnetic films (see above). We also see a strong increase in the amplitude of the fundamental FMR mode with the decrease in $w$. This increase is accompanied by a growth in the off-resonance characteristic inductance of the MSL loaded by the ferromagnetic film. For $w$=3 micron the inductance becomes equal to one of the MSL without a film on its top. The latter fact demonstrates that for small $w$ values the film becomes effectively insulating and its presence does not alter the transmission characteristics of the MSL off-resonance.

All these findings explain, why strong contribution of the microwave shielding effect to the broadband stripline FMR response of metallic ferromagnetic films has been observed in a number of experiments, but all existing experimental data (see e.g. [19-22]) do not show any indication of an eddy current impact on excitation of travelling spin waves in ferromagnetic metallic films. This happens just because for the FMR measurements wide macroscopic MLS are typically employed, but in order to efficiently excite travelling spin waves in a broad frequency range one needs much narrower transducers - 0.5µm<$w$<5 µm or so.

## 2. Numerical model

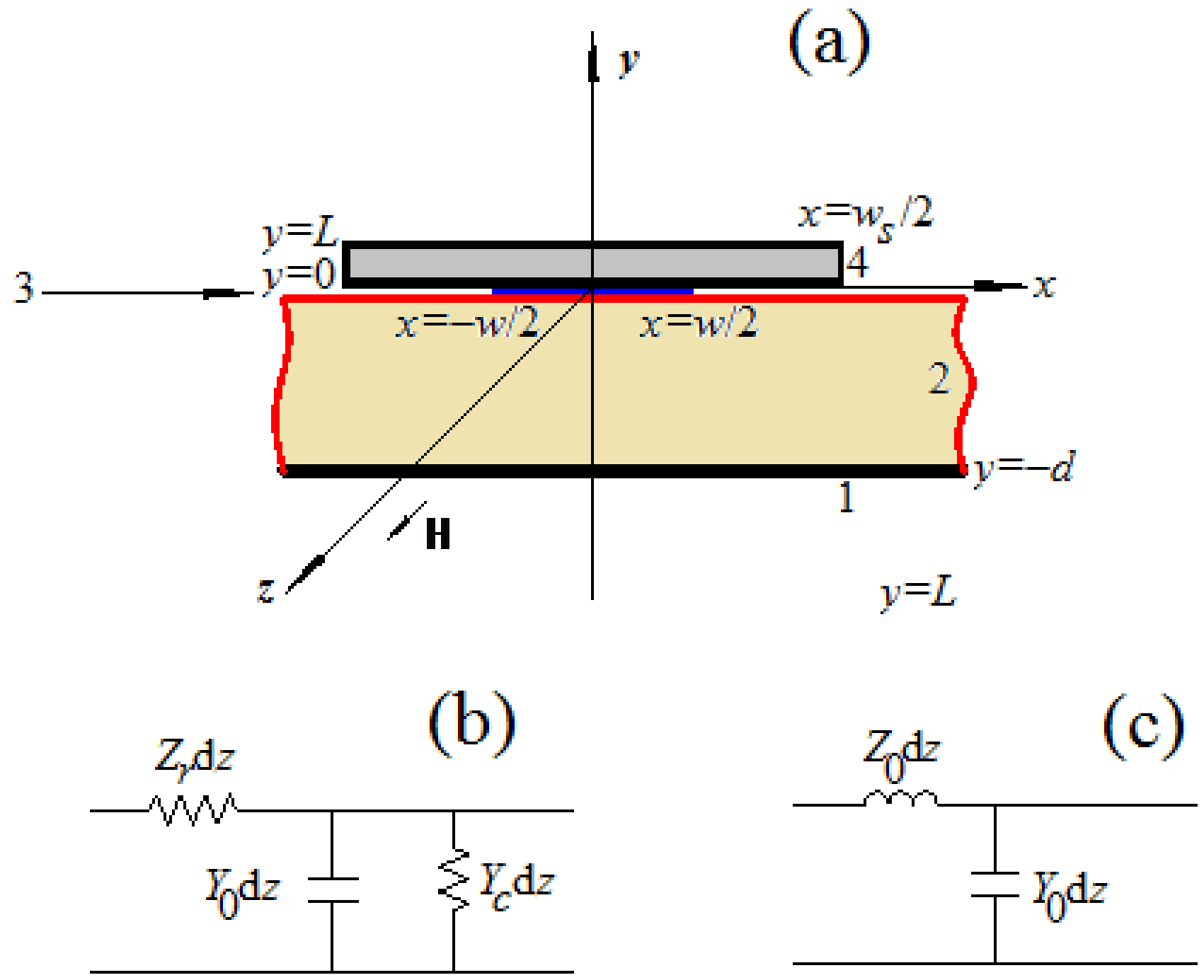

**Figure 1** (a) Sketch of the modelled geometry. 1: The ground plane of the microstrip line. 2: substrate of the microstrip line of thickness $d$. 3: infinitely thin strip of width $w$ carrying a microwave current. 4 ferromagnetic film of thickness $L$ in the direction $y$ and of width $w_s$ in the direction $x$. (b) Equivalent circuit of the microstrip line with the ferromagnetic film on its top (“loaded MSL”). The sketch shows one unit cell of the periodic chain. (c) Equivalent circuit of the microstrip line without film on its top ("unloaded MSL").

To solve the problem we use the numerical approach first suggested in [8]. We consider the model in which the *y*-axis is perpendicular to the surfaces of the conducting magnetic film (Fig. 1(a)). We assume that the film size $w_s$ in the direction *x* is larger than 2*w*. Because the microwave field of a microstrip line is well localized within the area of the width 2*w* below the strip of the microstrip line, this allows us to consider the sample as of an infinite width in the direction *x* in the following (but we will use the finite value of $w_s$ to post-process the results of the numerical simulations).

Thus, the film of thickness *L* is assumed to be continuous in the *x*- and *z*-directions. The external magnetic field $\mathbf{H} = H\mathbf{u}_z$ is applied in the positive direction of the *z*-axis. All the dynamic variables depend harmonically on time - $\exp(i\omega t)$, where $\omega$ is microwave frequency. In order to include the magnetization dynamics in the ferromagnetic layer into the model, we employ the linearized Landau-Lifshitz equation

$$i\omega t\mathbf{m} = -|\gamma|(\mathbf{m}\times\mathbf{H} + M_0\mathbf{u}_z\times\mathbf{h}_{eff}) \quad (1)$$

In (1) the dynamic magnetization vector **m** has only two non-vanishing components ($m_x$, $m_y$) that are perpendicular to the static magnetization $M_0\mathbf{u}_z$, where $M_0$ is saturation magnetization for the ferromagnetic film and $\mathbf{u}_z$ is a unit-vector in the direction *z*. The dynamic effective field $\mathbf{h}_{\mathrm{eff}}$ has two components: the exchange field $\mathbf{h}_{\mathrm{ex}}$ given by

$$\mathbf{h}_{exch} = \alpha(\partial^2/\partial x^2 + \partial^2/\partial y^2)\mathbf{m} \quad (2)$$

and the dynamic magnetic field **h**. In (2) the coefficient $\alpha$ is the exchange constant.

The calculation of the dynamic magnetic field in the presence of the eddy currents represents the main difficulty of the problem. For the exchange-free case ($\alpha$=0) a sophisticated analytical solution exists [23], but for $\alpha \neq 0$ constructing a numerical code is much more appropriate.

The dynamic magnetic field is solution of Maxwell equations:

$$\nabla\times\mathbf{h} = \sigma\mathbf{e}, \quad (3)$$

$$\nabla\times\mathbf{e} = -i\omega\mu_0(\mathbf{h}+\mathbf{m}), \quad (4)$$

$$\nabla\mathbf{h} = -\nabla\mathbf{m}, \quad (5)$$

where $\mu_0$ is the magnetic permittivity of vacuum and $\mathbf{h} = (h_x, h_y)$. Because the microwave magnetic field of MSL is quasi-static and also because the ferromagnetic film is metallic we neglected the term involving the electric permittivity on the right-hand side of Eq.(3) and take into account only the term involving the film conductivity $\sigma$.

We also assume that all dynamical variables do not depend on *z* (quasi-static approach to description of microwave transmission lines), therefore our problem is two-dimensional. The standard real-space methods face serious difficulties [1,18] even in two dimensions because of incompatibility of the length scales in Fig. 1(a) – compare *L* with *w* and *d*. To get around this problem and ultimately to accelerate the numerical solution we take advantage of the translational symmetry of the sample in the direction *x* (recall, we assume that the film is continuous in this direction). To this end we Fourier-transform Eqs.(3-5) and (1,2) with respect to *x*. This will allow us to remove the areas *y*<0 and *y*>*L* from the numerical model and thus to get around the scale incompatibility problem. Furthermore, equations for different Fourier wave numbers *k* separate. In principle, this allows solving the constructed numerical problem in parallel for large number of *k* values, and hence parallel computations.

To implement the Fourier transformation we assume that

$$\mathbf{m},\mathbf{h} = \int_{-\infty}^{\infty} \mathbf{m}_k, \mathbf{h}_k \exp(-iky)dy ,$$

where

$$\mathbf{m}_k, \mathbf{h}_k = 1/(2\pi)\int_{-\infty}^{\infty} \mathbf{m},\mathbf{h} \exp(iky)dy .$$

This procedure results in a system of equations:

$$\partial h_x / \partial y + ikh_y = -\sigma e_z \ , \qquad (6)$$

$$\partial e_z / \partial y = -i\omega\mu_0 (h_x + m_x) , \qquad (7)$$

$$ke_z = -\omega\mu_0 (h_y + m_y) , \qquad (8)$$

$$-ikh_x + \partial h_y / \partial y = -\partial m_y / \partial y + ikm_x \ . \qquad (9)$$

Here and at many places below we drop the subscript "*k*" to simplify notations.

We differentiate (6) and substitute (7) into the resulting differentiated equation and then make use of (9). This gives:

$$(\partial^2 / \partial y^2 - K^2)h_x - K^2 m_x - ik\partial m_y / \partial y = 0 , \qquad (10)$$

where $K^2 = k^2 + i\sigma\omega\mu_0$.

In this work Eqs. (10) and (9) will be solved numerically to produce the dynamic magnetic field. To obtain the numerical solution we need the electromagnetic boundary conditions relating the electromagnetic fields inside and outside the film. Furthermore, to accelerate the numerical computation it is useful to exclude the areas outside the film from the discrete model. This is possible, since an analytic solution for the space outside the film exists [3,8].

The microwave magnetic field outside the film is given by the same Eq.(10), but for $\sigma = \mathbf{m} = 0$. Let us first consider the area behind the film *y*>*L*. From (10) and the condition of the vanishing of the microwave magnetic field at *y*=+∞ one easily finds that outside the film (*y*>*L*)

$$h_y = -i\frac{|k|}{k}h_x , \ (11)$$

and at the film surface (*y*=*L*)

$$\frac{|k|}{k}(h_{yi} + m_{yi}) + ih_{xi} = 0 . \qquad (12)$$

In this expression the subscript "*i*" indicates that these field components are taken at the film surface from *inside* the film. Eq.(12) represents the electromagnetic boundary condition at *y*=*L* which excludes the area *y*>*L* from consideration.

A similar boundary condition can be also obtained for the area in front of the film. This area contains the strip and the ground plane of the MSL. We model the strip as an infinitely thin sheet of microwave current $I\mathbf{u}_z$ with uniform current density $j = I / w$ (Fig. 1). The width of the sheet along the *x*-axis is *w*. The sheet is infinite in the *z* direction to ensure continuity of the current. The Fourier image of the current density reads:

$$j_k = \frac{I}{2\pi}\frac{\sin(kw/2)}{kw/2} \quad (13).$$

For simplicity we place the current directly on the surface of the film (*y*=0). This allows one employing the electro-dynamic boundary condition which includes a surface current. This boundary condition reads (here we write it down in the Fourier space):

$$h_{xi} = h_{xe} - j_k \ , \ (14)$$

where the subscript "*e*" denotes the area outside the film.

The MSL ground plain is modelled as a surface of a metal with infinite conductivity ("ideal metal"). It is located at y=–*d*. At the ideal-metal surface $e_z$=0. By solving (10) for $-d<y<0$ analytically together with this boundary condition and the boundary condition (14) one obtains a boundary condition at *y*=0 as follows:

$$(h_{yi} + m_{yi})\coth(|k|d) - i\frac{|k|}{k}h_{xi} = i\frac{|k|}{k}j_k . \quad (15)$$

One sees that Eq.(15) includes only components inside the film and hence excludes the area in front of the film from consideration in the numerical model.

The system of equations (1), (10), (9), (12) and (15) will be solved numerically to obtain the dependence $\mathbf{m}_k(y)$. This system is inhomogeneous, since Eq.(15) contains an excitation term given by its right-hand side. In the next section we will give details of the numerical method. Before we proceed to the next section we need to introduce equations which will be used to post-process results of the numerical calculation.

In particular, we will be interested in the complex surface impedance of the film $Z_r$. This quantity is a measure of the microwave magnetic absorption by the film [10]. It can be defined as follows:

$$Z_r = -\frac{U}{I}, \quad (16)$$

where the linear voltage *U* (measured in V/m) is the mean value of the total electric field induced at the surface of the strip of MSL:

$$U = \frac{1}{w}\int_{-w/2}^{w/2} e_z(y=0)dx = \int_{-\infty}^{\infty} e_{zk}(y=0)\frac{\sin(kw/2)}{kw/2}dk . \ (17)$$

Since no electric field is externally applied to the MSL in our model, the source of this voltage is the dynamic fields of the film. Hence the negative sign in Eq.(16).

The electric field $e_{zk}$ can be easily obtained with Eq.(6), once $\mathbf{m}_k$ and $\mathbf{h}_k$ have been calculated numerically.

Once $Z_r$ has been computed, it is a straightforward procedure to calculate S21 [3]. We make use of the equivalent circuits for a microstrip line with the film on its top ("loaded MSL", Fig. 1(b)) and for an MSL without the film ("unloaded MSL", Fig. 1(c)). The characteristic impedance for the loaded MSL is

$$Z_f = \sqrt{Z_r / (Y_0 + Y_c)}\,, \quad (18)$$

where $Y_0$ is the intrinsic (i.e. in the absence of the sample) parallel capacitive conductance of MLS and $Y_c$ is the parallel capacitive conductance due to the electric shielding effect (see introduction). It is obvious that far off FMR $Z_r$ should become purely imaginary and reduce to the in-series inductive impedance of MSL with a thin film on its top. Furthermore, for $\sigma$=0 the latter quantity should further reduce to the intrinsic in-series inductive impedance of MSL $Z_0$ (Fig. 1(c )). Indeed, our numerical calculations will show that off resonance $Z_f(\sigma=0)$ converges to $Z_0$. This justifies, why we do not include $Z_0$ into (18).

$Z_0$ can be calculated using the standard formalism for microstrip transmission lines [24]. For $Z_r$=$Z_0$ and $Y_c = 0$ Eq.(18) reduces to the expression for the intrinsic characteristic impedance of MSL:

$$Z_c = \sqrt{Z_0 / Y_0} \quad (19).$$

The intrinsic propagation constant for MSL is given by the formula as follows:

$$\gamma_0 = i\omega\sqrt{\varepsilon_0 \varepsilon_{eff} \mu_0}\,, \qquad (20)$$

where $\varepsilon_{eff}$ is the effective electric permittivity of MSL

$$\varepsilon_{eff} = \frac{\varepsilon_s + 1}{2} + \frac{\varepsilon_s - 1}{2}\frac{1}{\sqrt{1 + 12d / w}}\,, \quad (21)$$

$\varepsilon_s$ is the relative dielectric permittivity of the MLS substrate and $\varepsilon_0$ is the dielectric permittivity of vacuum. The expression for $Z_c$ can be found in any textbook on microwaves (see e.g. [24]), therefore we do not show it here. Importantly, once $Z_c$ and $\gamma_0$ have been calculated, $Z_0$ and $Y_0$ can be obtained by solving (19) together with

$$\gamma_0 = \sqrt{Z_0 Y_0} \quad (22).$$

As shown in [3], the complex transmission coefficient of a section of MSL with the film on its top is given by

$$S21 = \frac{\Gamma^2 - 1}{\Gamma^2 \exp(-\gamma_f l_s) - \exp(\gamma_f l_s)}\,, \qquad (23)$$

where $l_s$ is the length of the film in the direction $z$, $\Gamma$ is the complex reflection coefficient from the front edge of the section of MSL covered by the film

$$\Gamma = \frac{Z_f - Z_c}{Z_f + Z_c} \quad (24)$$

and

$$\gamma_f = \sqrt{(Z_r(Y_0 + Y_c)} \quad (25)$$

is the propagation constant for MSL loaded by the film. To make use of (23) we also need $Y_c$. Derivation of a suitable expression for $Y_c$ is given in Appendix 1.

## 3. The discrete model

To solve the system of equations (1), (10) and (9) numerically we use the finite difference approach. These equations are differential equations of only one spatial variable – $y$. Accordingly, we employ a one-dimensional mesh. The number of mesh points is $n$ and they all are located inside the film, i.e. between $y$=0 and $y$=$L$. The equations are discretized by using the standard 3-point formula for numerical calculation of derivatives. The boundary conditions (12) and (15) are included into the discrete version of (10) and (9) for the mesh points at the film surfaces $y=\Delta/2$ and $y=L-\Delta/2$, where $\Delta=L/n$ is the mesh step. In the same way we discretize the exchange operator (2) and include the exchange boundary conditions for dynamic magnetization into it. For simplicity we assume that **m** is unpinned at both film surfaces $y$=0 and $y$= $L$:

$$\partial \mathbf{m} / \partial y = 0 \,. \quad (26)$$

The method of inclusion of the boundary conditions into the differential operators was explained in Appendix to [3], therefore we do not repeat its description here.

The dicretization of (10) simultaneously with (9) results in a vector-matrix equation in the form

$$\hat{H} \mid h_k > + \hat{M} \mid m_k >= \hat{H}_{mw} \mid h_{mwk} > . \qquad (27)$$

Similarly, Eqs.(1) and (2) are cast in the form

$$\mid h_k >= \hat{G} \mid m_k > . \qquad (28)$$

In these expressions $\mid h_k >$ and $\mid m_k >$ are column vectors containing values of $x$ and $y$ components of $\mathbf{h}_k$ and $\mathbf{m}_k$ respectively at the mesh points. The vector $\mid h_{mwk} >$ contains only one non-vanishing element. It is the $x$-component at $y=\Delta/2$. It is equal to $j_k$. This is consistent with Eq.(14) which relates the $x$-component of the dynamic field at the film surface to the current density in the microstrip.

Combining (27) and (28) one obtains

$$\hat{H}\hat{G} | m_k > + \hat{M} | m_k > = \hat{H}_{mw} | h_{mwk} > . \quad (29)$$

In this work we solve Eq.(29) for $| m_k >$ numerically for $I$=1A and given values of $H$, and $\omega$ using the standard numerical methods of linear algebra. The result is then utilized to compute $e_{zk}(y=0)$ with (28) and (6). This is done for a large number of $k$-values for which $j_k$ (Eq.(13)) is non-negligible. Then we use the "modified" inverse Fourier transform (17) in order to calculate $U$. $Z_r$ and S21 are then obtained with (16) and (23) respectively.

**4. Discussion**

This algorithm has been implemented as a MathCAD worksheet. Therefore the computation time is rather long – 12 hours for one program run during which $e_{zk}(y=0)$ is computed for 4096 $k$-values and 80 $H$-values for a given frequency and $n$=100. However, our Fourier-space formulation of the problem should make it perfectly suitable for parallel computing. Indeed, using multiple processors large number of independent solutions of (29) for different values of $k$ may be obtained in parallel. This should drastically decrease the computation time. Furthermore, rewriting the code in a language which is more appropriate for lengthy computations than the programming tool built in into MathCAD software will make computations much faster, even without parallelization.

Several tests of the produced code were run in order to make sure that the software delivers correct results. The first test was to solve (27) assuming $| m_k > = 0$ and $w$ is large ($w$=1.5mm). For $L$=100nm and $\sigma$=4.5x$10^6$Sm/m which is the typical conductivity of Permalloy ($Ni_{80}Fe_{20}$), we found that $h_x(y=0,x=0)=j$ and $h_x(y=L,x=0)=0$. This is consistent with previous theories of the microwave shielding for thin non-magnetic metallic films ([3,18]). For a $\sigma$–value close to zero we found that $Z_r=Z_0$. The latter (i.e. $Z_0$) was calculated analytically using (20) and (22) and the standard expression for the characteristic impedance of MSL [24].

Also, for $\alpha$=0 an analytic solution of (1), (9) and (10) exists. It is similar to one in Ref. [23] and will be reported elsewhere. Our solution of the full system (29) (i.e. for $| m_k > \neq 0$) was checked against this analytic solution and showed excellent agreement.

Below we demonstrate some important results of our computations. Figure 2 displays $Z_r(H)$ for different values of $w$. The frequency is 15 GHz and the Permalloy film thickness is either 50nm or 100nm. Since we keep the frequency constant and vary the applied field, this simulation mimics the conditions of the field-resolved FMR experiment.

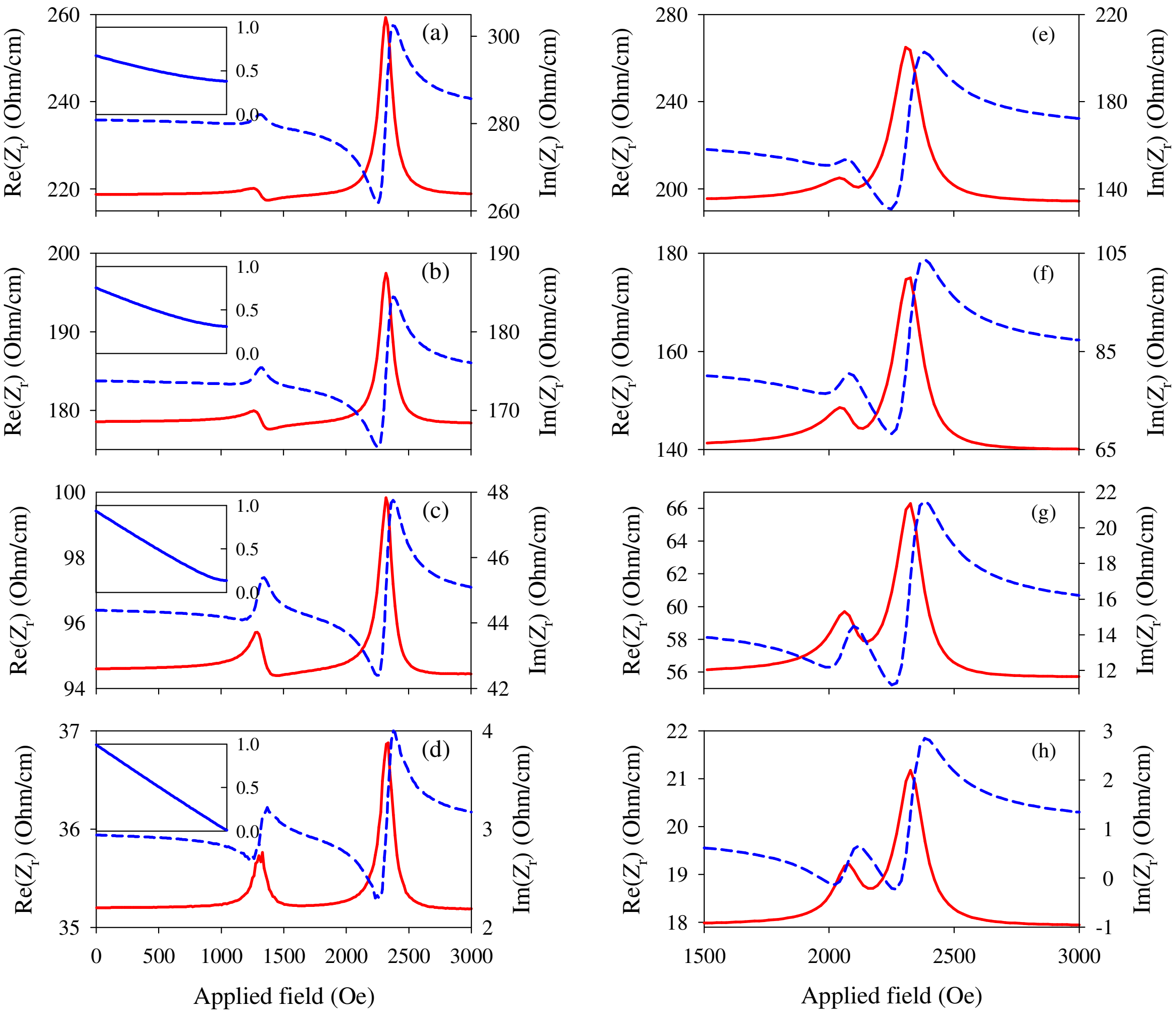

Fig. 2. Surface impedance of the film for different values of the strip width $w$. (a)-(d): film thickness $L$=50nm. (e)-(h): $L$=100nm. (a) and (e): strip width $w$=50μm; (b) and (f): $w$=100μm; (c) and (g): $w$=350μm; (d) and (h): $w$=1500μm.

Insets to Panels (a)-(d) show respective profiles $h_x(y)$ of the microwave magnetic field across the film thickness at the symmetry plane of the strip $x$=0. They are calculated for the maximums of the resonance line for the fundamental mode. The horizontal axes of the graphs in the insets are not labelled, in order not to overload the panels. Left edges of the graphs correspond to $y$=0 and the right edges to $y$=$L$.

Parameters of calculation. Microwave frequency: 15 GHz; saturation magnetization $\mu_0 M_0$=1T; magnetic loss parameter for the ferromagnetic film $\Delta H$=6mT. The microstrip substrate thickness $d$ and its electric permittivity $\varepsilon_s$ are chosen such that the intrinsic characteristic impedance of the microstrip line $Z_c$=50 Ohm for all panels: (a) and (e): $d$=100μm, $\varepsilon_s$ =16; (b) and (f): $d$=100μm, $\varepsilon_s$ =9; (c) and (g): $d$=160μm, $\varepsilon_s$=3.55; (d) and (h): $d$=700μm, $\varepsilon_s$=3.55. Film conductivity $\sigma$=4.5x$10^6$Sm/m. Solid lines: Re($Z_r$); dashed lines: Im($Z_r$).

The first observation from this figure is that the amplitude of first higher-order standing spin wave mode ($1^{st}$ SSWM, located at 1290 Oe for $L$=50nm and at 2060 Oe for 100nm) is significant, although the dynamic magnetization vector is unpinned at both film surfaces. This mode is excited due to the eddy-current contribution to FMR response.  For $w$=1.5mm the relative amplitude of this mode is the same as given by the 1-dimensional theory from [3]. It becomes almost negligible with the decrease in $w$. The latter result is the main funding of this work.

As previously shown in [18], the microwave shielding effect gradually disappears with a decrease in $w$. This is confirmed by the present calculation for the ferromagnetic films (see insets to Fig. 2). As Fig. 2 demonstrates, this leads to suppression of excitation of the 1$^{st}$ SSWM. The possibility of the stripline FMR to excite the 1$^{st}$ SSWM is very important for characterization of magnetic films, since its frequency and field position with respect to the fundamental mode carries information about the value of the exchange constant for the material (see Eq.(2)). The cavity FMR is unable to excite this mode, unless significant pinning of magnetisation is present at one of the film surfaces. This is because the eddy currents vanish in the conditions of the cavity FMR for symmetry reasons (see e.g. [17]). As follows from Fig. 2, the stripline FMR has an important advantage from this point of view, provided one uses wide striplines or CPWs.

The next observation from this figure is that the amplitude of the fundamental FMR peak (the larger peak in each panel) grows significantly with the decrease in $w$. This fact is well-known to experimentalists. Therefore they tend to use narrow striplines (100 to 300nm wide) for their measurements (see e.g. [25]). However, in this way, they unintentionally decrease the response of the 1$^{st}$ SSWM, as it is clear from this figure.

From Fig. 3 one also sees that the off-resonance values of Re($Z_r$) and Im($Z_r$) grow with the decrease in $w$. This effect can be also attributed to the decrease in the microwave shielding effect. Loading of a wide MSL by a metallic film decreases the in-series inductive impedance of MSL with respect to $Z_0$. The decrease is larger for wider stripes. For instance, for $w$=1.5mm $Z_0$=270 Ohm/cm, but the off-resonance value of Im($Z_r$) for $L$=100nm and $w$=1.5mm in Fig. 2 is about 1 Ohm/cm.

For $w$=3µm, however, the Im($Z_r$) off resonance is 540 Ohm/cm (as measured at $H$=800 Oe, Fig. 3). This is close to $Z_0$ for this value of $w$ - 620 Ohm/cm. To demonstrate the disappearance of shielding, in Fig. 3 we also show the result of calculation for a negligible value of film conductivity (5 Sm/m). One sees that the two results are very close to each other. This shows that the film becomes effectively insulating for small stripline widths.

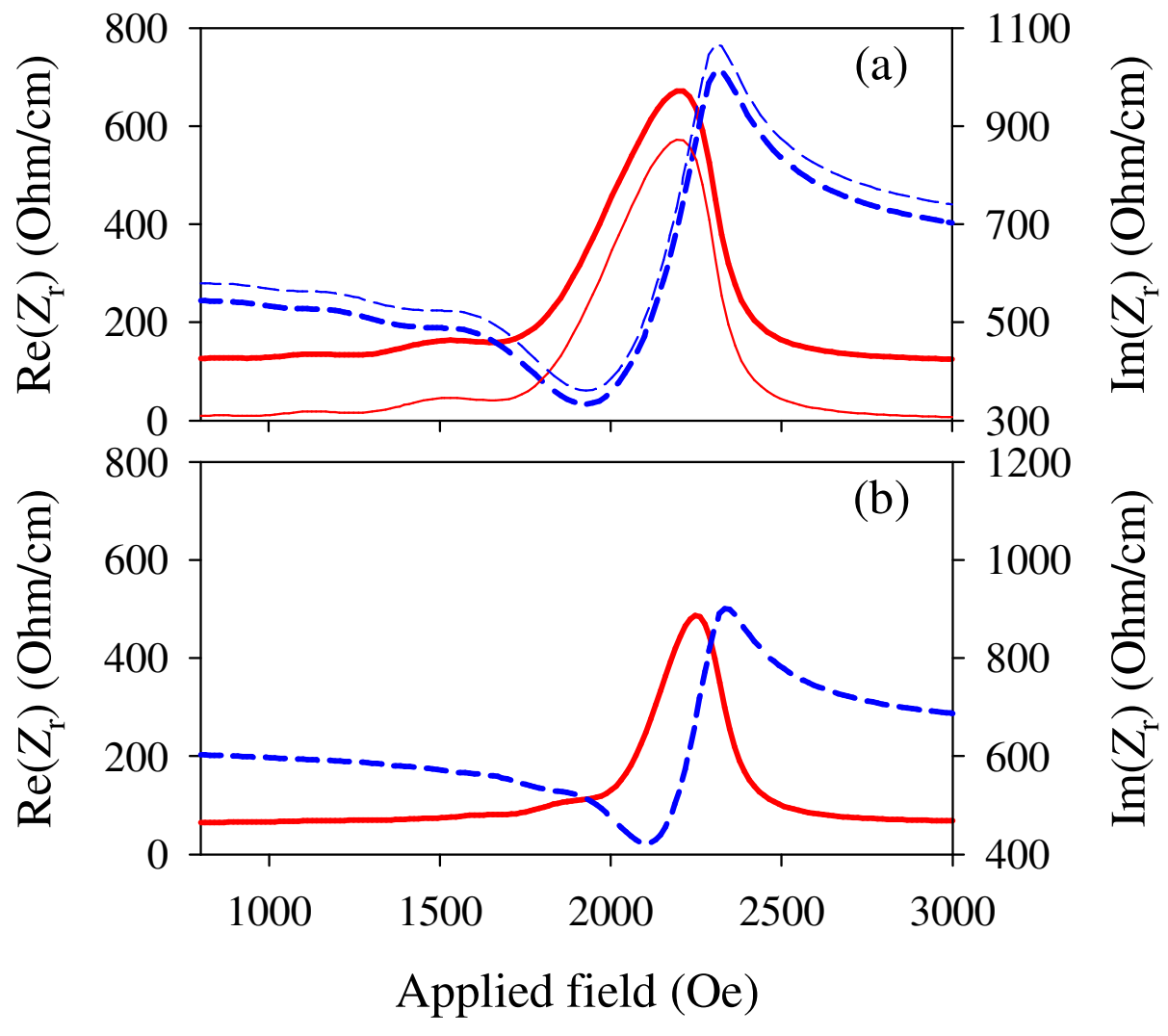

Fig. 3. Surface impedance for a narrow strip - $w$=3µm. (a) Film thickness $L$=100nm. (b) $L$=50nm. $d$=10µm and $\varepsilon_s$ =16 for both panels. All other calculation parameters are the same as for Fig. 2. (The unrealistically small value of $d$ was chosen in order to obtain $Z_0$=50 Ohm, and hence to allow comparison of Figs. 2 and 3.) Thick lines: $\sigma$=4.5x10$^6$Sm/m; thin lines: $\sigma$=5 Sm/m. Solid lines: Re($Z_r$); dashed lines: Im($Z_r$).

Another important observation from Fig. 3 is the total disappearance of the 1st SSWM peak and appearance of small waviness of Re($Z_r$) for smaller applied fields. The total absence of the SSWM peak is in agreement with the idea of the complete disappearance of the shielding effect. The appearance of the waviness, however, does not relate to the film conductivity, since it is the same for both conducting and insulating films in the upper panel of Fig. 3. The waviness has to be considered in conjunction with significant broadening of the fundamental resonance peak seen in Fig. 3. This broadening is due to contribution of travelling spin waves to the stripline FMR response [26]. For $L$=100nm and $w$=3micron the width of the strip is significantly smaller than the free propagation path of spin waves – tens of micron. The spin waves are efficiently excited by this MSL "antenna" and leave the area of localization of the driving field. They carry energy away from the antenna which is seen as additional resonance losses leading to the broadening of the resonance line.

An alternative explanation to the spin wave contribution is as follows. $Z_r(k)$ (where $k$ has now sense of spin wave wave number) scales roughly as square of $\frac{\sin(kw/2)}{kw/2}$. This is seen from Eqs.(13) and (17). The first lobe of this function is located between $k$=0 and $k=2\pi/w$. Most of microwave power goes into energy of spin waves in this wave number range. This wave number range corresponds to an applied-field range roughly equal to $V_g 2\pi/(|\gamma|w)$, where $V_g$ is the spin wave group velocity. When the intrinsic resonance linewidth (intrinsic magnetic loss parameter) for the material $\Delta H$ is larger than $V_g 2\pi/(|\gamma|w)$, the spin wave contribution to the stripline FMR linewidth is negligible. On the contrary, when $\Delta H<V_g 2\pi/(|\gamma|w)$, non-negligible the spin wave contribution will exist. The group velocity of the Damon-Eschbach spin wave [27] for $L$=50 nm is twice smaller than for $L$=100nm. Therefore the peak width in Fig. 3 for $L$=50nm is narrower by approximately two times.

The waviness in Fig. 3 is reflection of the fact that $Z_r(k)\sim\left(\frac{\sin(kw/2)}{kw/2}\right)^2$. Therefore, because for small spin wave wave numbers the spin wave dispersion law is practically linear, the shape of $Z_r(H)$ is close to the shape of the $\sin^2(x)/x^2$ function.

The very small difference in the shapes and amplitudes of the results for $\sigma$=0 and $\sigma\neq0$ in Fig. 3(a) explains why conductivity effects have never been observed in the experiments on excitation of travelling spin waves in ferromagnetic metals with stripline antennas [19-21]. Indeed, in order to excite spin waves in a noticeable frequency or wave number range, very narrow antennas are needed (see Fig. 3). For the narrow antennas the conductivity effects are negligible, as follows from above.

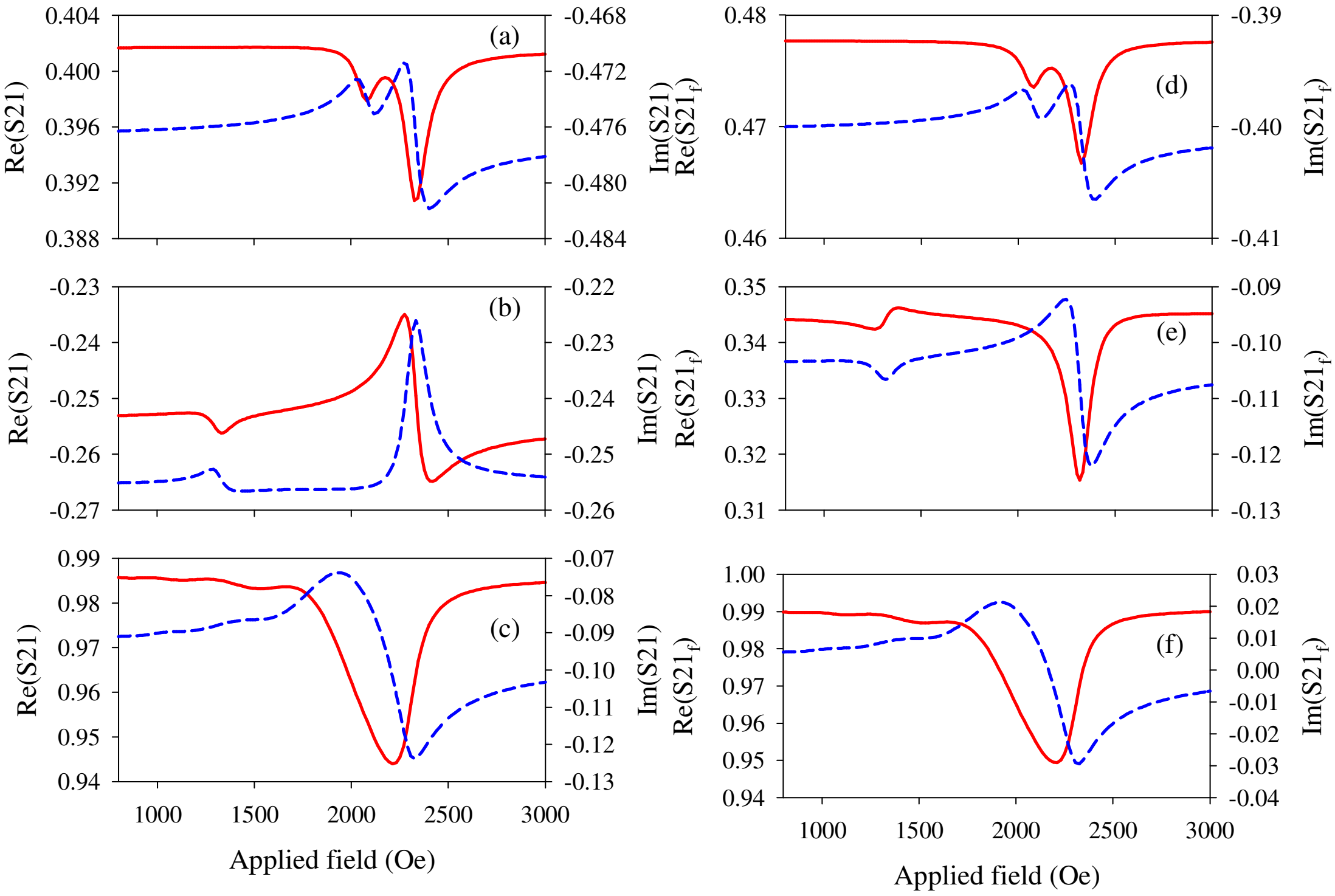

Fig. 4. (a-c): Unprocessed S21 data. (d-f): processed S21 data. (a) and (d): $L$=100nm, $w$=1500μm; (b) and (e): $L$=50nm and w=100μm; (c) and (f): $L$=100nm, $w$=3μm. All other parameters of calculation are the same as for Figs. 2 and 3. Solid lines: Re(S21); dashed lines: Im(S21). (d) $\phi$=–110°. (e) $\phi$=–60°. (g) $\phi$=0°.

In Fig. 4(a-c) we demonstrate the S21($H$) dependence calculated with the data from Figs. 2(b), 2(h) and 3(a). One sees that we obtain realistic values of S21 and the shapes of its real and imaginary parts are very reasonable. Interestingly, similar to Fig. 2(b) in Fig. 4(b) one observes different shapes for the fundamental and the 1st SSWM peaks. The fundamental peak in Re(S21) has a shape close to Lorentzian, which is the characteristic shape for the imaginary part of the complex microwave magnetic permeability [28], but the response of the 1st SSWM has a shape characteristic to the real part of the complex permeability. This effect of difference in the peak shapes is often seen in the experiment (see e.g. Fig. 2(c) in [4]).

Another observation from Fig 4(a-c) is that the shapes of Re(S21) and Im(S21) change from panel to panel. However, if one divides the "raw S21 data" from Fig. 4(a-c) by $S21_0=S21(Z_f=Z_0,\gamma_f=\gamma_0)\exp(-i\phi)$, where $\phi$ is phase correction, one obtains traces of "standard" shapes, as seen in panels (d-f). The latter panels display $S21_f=S21/S21_0$ for values of $\phi$ which were carefully chosen to produce the most symmetric (anti-symmetric) shapes of Re(S21) (Im(S21)) respectively. Previously a similar procedure was employed to process experimental data [4]. Traces of S21 taken without a sample on top of the microstrip line were used as $S21_0$ in that publication. This procedure also resulted in traces of the "standard" shape (see Fig. 2(a-d) in that paper).

## Conclusion

In this work we constructed a two-dimensional numerical model for calculation of the stripline ferromagnetic resonance response of metallic ferromagnetic films. We also conducted numerical calculations by using this software. The calculations demonstrated that the eddy current contribution to the FMR response decreases with a decrease in the stripline width. The most important manifestations of the conductivity (eddy current) effect are (1) excitation of the higher-order standing spin waves across the film thickness in the materials for which the standing spin wave peaks would be absent in cavity FMR measurements and (2) strong dependence of the off-resonance series conductance of the stripline on the stripline width.

The model is suitable for description of both stripline ferromagnetic resonance experiments and experiments on excitation of travelling spin waves. It also demonstrates gradual transition from the former regime to the latter one with a decrease in the width of the stripline. Whereas the contribution of the eddy currents to the stripline FMR response can be very significant, because wide striplines (100nm+) are conventionally used for the FMR measurements, it is negligible in the case of excitation of spin waves, just because very narrow stripline transducers (0.5-5micron wide) are required in order to excite spin waves in metallic ferromagnetic films in a noticeable frequency/applied field range.

For simplicity we considered magnetization dynamics in uniform ferromagnetic films. However, it is known that for films lacking inversion symmetry in the direction of the film thickness the impact of eddy currents on the stripline FMR response may be even more pronounced. Examples are bi-layer films [4] and single-layer metallic films with unequal strengths of magnetization pinning at the film surfaces [17]. The stripline FMR response of these films strongly depends on the film orientation (e.g. layer ordering) with respect to the stripline [5]. The results of the present study indicate that for these materials one may also expect a gradual decrease in the strength of the eddy current effects with a decrease in the stripline width and hence a decrease in the difference in the FMR responses for the two possible film orientations with respect to the stripline.

**Acknowledgment**
Support by the Australian Research Council, the University of Western Australia (UWA) and the UWA's Faculty of Science is acknowledged. Z.L. acknowledges the student exchange program between USTC and UWA. We also thank I.S. Maksymov for fruitful discussions.

**Appendix: Derivation of an expression for the parallel capacitive conductance $Y_c$.**

Here we derive an expression which allows estimation of $Y_c$ (Fig. 1(b)). This parameter is a measure of the strength of the electric shielding [1]. We calculate $Y_c$ based on the equivalent circuit in Fig. A1. We assume that the capacitance between the strip and the film is infinite. This is a valid assumption, since the film is located directly on the strip surface (Fig. 1(a)). Due to the electric shielding effect a microwave current flows along the film in the direction $x$ (Figs. 1(a) and A1) and then from the film to the stripline ground plane due to the distributed capacitance between the film and the stripline $C\Delta x\Delta z = \varepsilon_s\varepsilon_0 \Delta x\Delta z / d$ (Fig A1). The resistance of a unit area $\Delta x\Delta z$ of the film $R\Delta x=\Delta x/(\sigma L\Delta z)$. The total length of the $R$-$C$ chain is $(w_s-w)/2$ and it is terminated by an open end (the edge of the film). There are two such "chains": to the right and to the left from the strip.

The input impedance $Z_{in}$ of each chain can be calculated by using the same formalism of Eqs. (19-25). The in-series linear impedance $Z=R\Delta x$ and the linear parallel conductance is $Y=i\omega C\Delta x$. The characteristic impedance of a chain is then $Z_{cc} = \sqrt{Z/Y} = \Delta z \sqrt{d/(i\varepsilon_s \varepsilon_0 \sigma L \omega)}$. The propagation constant is $\gamma_{cc} = \sqrt{ZY} = \sqrt{i\varepsilon_s \varepsilon_0 \omega/(\sigma L d)}$. For a chain terminated by an open end, $Z_{in}$ is given by a formula as follows: $Z_{in} = Z_{cc} \coth(\gamma(w_s - w)/2)$. Then, taking into account that there are two identical chains (Fig. A1), we obtain the final formula

$$Y_c = 2\Delta z / Z_{in} = 2\sqrt{(i\varepsilon_s \varepsilon_0 \sigma L \omega)/d}\ \tanh[\sqrt{i\varepsilon_s \varepsilon_0 \omega/(\sigma L d)}(w_s - w)/2]. \quad \text{(A1).}$$

Calculations show that $Y_c$ is usually significantly smaller than $Y_0$. For instance, for $f$=15GHz, $L$=100nm, $w$=1.5mm, $w_s$=5mm, $d$=0.7mm and $\varepsilon_s$=3.55 one obtains $Y_0$=10.1$i$ Ohm/m and $Y_c$=6.1x10$^{-4}$(1+$i$).

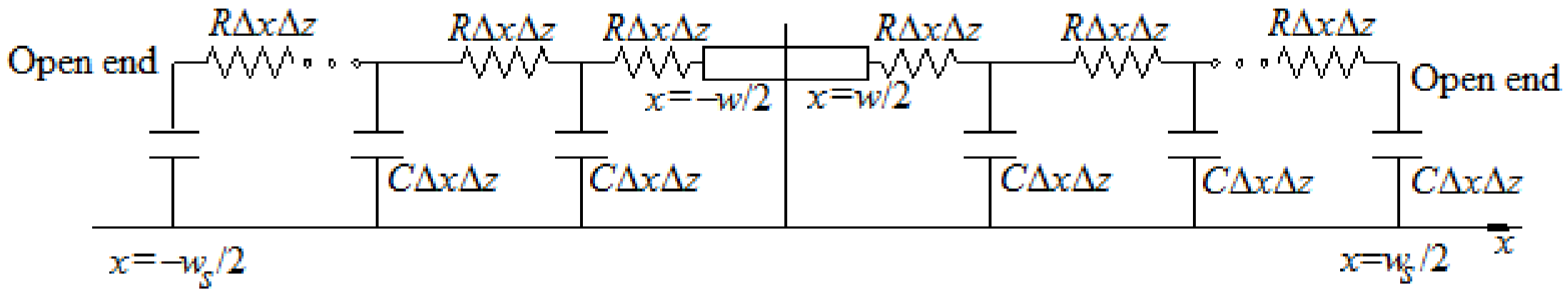

Fig. A1. Equivalent circuit for estimating $Y_c$.